---

# A Novel ANROA Based Control Approach for Grid-Tied Multi-Functional Solar Energy Conversion System

Dinanath Prasad[1,2], Narendra Kumar[1], Rakhi Sharma[3], Hasmat Malik[4], Fausto Pedro Garcia Márquez[5], Jesús María Pinar- Pérez[6,*]

[1] Delhi Technological University, Electrical engineering Dept., Delhi, India. Dinanath Prasad: prasaddinanath@akgec.ac.in. Narendra Kumar: dnk_1963@yahoo.com.
[2] Ajay Kumar Garg Engineering College, Ghaziabad, India.
[3] IGNOU, Delhi, India, rakhisharma@ignou.ac.in.
[4] Division of Electrical Power Engineering, School of Electrical Engineering, Faculty of Engineering, Universiti Teknologi Malaysia (UTM), Johor Bahru 81310, Malaysia, hasmat.malik@gmail.com.
[5] Ingenium Research Group, Universidad Castilla-La Mancha, 13071 Ciudad Real, Spain, faustopedro.garcia@uclm.es.
[6] CUNEF Universidad, Leonardo Prieto Castro 2, 28040 Madrid, Spain, jesusmaria.pinar@cunef.edu.

**Abstract.** An adaptive control approach for a three-phase grid-interfaced solar photovoltaic system based on the new Neuro-Fuzzy Inference System with Rain Optimization Algorithm (ANROA) methodology is proposed and discussed in this manuscript. This method incorporates an Adaptive Neuro-fuzzy Inference System (ANFIS) with a Rain Optimization Algorithm (ROA). The ANFIS controller has excellent maximum tracking capability because it includes features of both neural and fuzzy techniques. The ROA technique is in charge of controlling the voltage source converter switching. Avoiding power quality problems including voltage fluctuations, harmonics, and flickers as well as unbalanced loads and reactive power usage is the major goal. Besides, the proposed method performs at zero voltage regulation and unity power factor modes. The suggested control approach has been modeled and simulated, and its performance has been assessed using existing alternative methods. A statistical analysis of proposed and existing techniques has been also presented and discussed. The results of the simulations demonstrate that, when compared to alternative approaches, the suggested strategy may properly and effectively identify the best global solutions. Furthermore, the system's robustness has been studied by using MATLAB/SIMULINK environment and experimentally by Field Programmable Gate Arrays Controller (FPGA)-based Hardware-in-Loop (HLL).

**Nomenclature:**

| | |
|---|---|
| ANF | Adaptive notch filter |
| ANFIS | Artificial Neuro-fuzzy inference system |
| ANROA | Neuro-fuzzy inference system with Rain optimization algorithm |
| BNC | Bayonet Neil-Concelman |
| DSOGI | Damped second-order generalized integrator |
| Du-SOGI | Dual second-order generalized integrator |
| FC | Fundamental component |
| FPGA | Field programmable gate arrays controller |
| GMPP | Global maximal power point |
| HLL | Hardware-in-loop |
| HPO | Human psychology optimization |
| IRPT | Instantaneous reactive power theory |
| LLMS | Leaky Least Mean Squares |
| MFOGI | Multilayer Fifth Order Generalized Integrator |
| MPPT | Maximal power point track |
| MSO | Mixed-signal oscilloscope |
| MSOGI | Modified second order generalised integrator |
| P&O | Perturb and observe |
| PCC | Point at common coupling |
| PSO | Particle swarn optimization |
| PI | Proportionate Integral |
| PV | Photovoltaic |
| ROA | Rain optimization algorithm |
| SECS | Solar energy conversion systems |
| SOGI | Second-order generalised integrator |
| SO-SOGI | Second-order SOGI |
| SPV | Standalone solar photovoltaic |
| SRFT | Synchronous reference frame theory |
| THD | Total harmonic distortion |
| VSC | Voltage source converter |

## 1. Introduction

The matrices used to create some regional grid characteristics are commonly feeble, untrustworthy, and have a poor quality of power (Eloy-García, Vasquez, & Guerrero, 2013; Zeng, Li, Tang, Yang, & Zhao, 2016). Accordingly, a combination of sustainable power sources to the grid network generates additional difficulties, aside from common challenges (Impram, Nese, & Oral, 2020). The grid can serve as an energy buffer in a grid-associated architecture, and energy generated by a solar photovoltaic system can be fed into the grid (Goud & Gupta, 2019). Solar photovoltaic (PV) energy focuses on the different PV system advancements because of damaging effects of continuous and extensive utilization of fossil fuels. For a photovoltaic array, Maximal Power Point (MPP) is an inimitable operating point where greatest power is produced. Nonetheless, due to irregular load or non-linear load, network voltage is affected by power quality-related issues such as under-voltage, DC balance, over-voltage, and harmonics (Ali, Christofides, Hadjidemetriou, & Kyriakides, 2019). Moreover, due to the non-linear nature of PV, maximum power point tracking techniques are needed to extract the most power from a PV array. Different MPPT methods have been proposed, including perturb and observe (P&O) (Esram, and Chapman, 2007), incremental conductance, and fuzzy-based methods (Saidi, Naceur, Mahjoub, & Bhutto, 2021). Managing photovoltaic (PV) power in a grid and its related issues becomes a troublesome task (Gawhade & Ojha, 2021). In the literature, various solar-fed grid-tied topologies with improved power quality have been discussed (Miska, Abdelaziz, & Akella, 2021). Solar energy conversion systems (SECS) have been established to help support the grid during peak load hours and in distant areas (Yang, Blaabjerg, Wang, & Simões, 2016). However, a SECS generates DC power. Connecting DC power to an AC network needs viable geographies and control to create a general framework of energy effectiveness (Agarwal, Hussain, &

Singh, 2017). Additionally, considering a microgrid, these problems are more prevailing due to nonlinear loading, sudden load variations, etc., that change as distortions are formed in the network voltage, frequency variations, etc. (Kumar, Hussain, Singh, & Panigrahi, 2017).

The presented SPV framework was a photovoltaic solar energy conversion framework interconnected with multi-working grids that alongside DC power change as SPV to AC grid was fit for reactive power compensation, elimination of harmonics currents, and load adjustment at three-stage AC distribution network. The implementation and trial outcomes check the legitimacy of the introduced approach and affirm its attractive transient and consistent state characteristics. The grid current THD (Total Harmonics Distortion) was discovered well in IEEE-519 standard considerably at nonlinear loads. Leaky Least Mean Squares (LLMS) depend on the control approach for a grid-connected three-phase single-stage solar energy conversion system (SECS). The principal goal of this framework was to eliminate the problems of power quality like voltage fluctuations and flickers, harmonics, imbalances on loads, reactive power requirements, etc. Also, this framework works at zero voltage regulation and unity power factor modes. In this examination, SECS utilizes solitary stage geography controlled via maximal power point tracking strategy and removing load currents, a crucial part, an adaptive neural network control technique based on LLMS was used. The framework was demonstrated, structured, and implemented in MATLAB utilizing simpower framework, and test approval were completed on a created model in the laboratory (Agarwal, Hussain & Singh, 2016). Additionally, a single input two phase grid solar photovoltaic (PV) associated stage partially shaded adaptive control technique dependent on Multilayer Fifth Order Generalized Integrator (MFOGI) and fuzzy tuned Proportionate Integral (PI) controller was used to track the global maximal power point (GMPP) using new human psychology optimization (HPO). MFOGI was utilized to extract the essential segment of grid voltage, in any event, while the grid voltage was categorized by overvoltage, under voltage, severe distortion of harmonics, variations of the frequency, direct current (DC) balance so on, in a broad range. This control approach was demonstrated and implemented in MATLAB (Kumar, Hussain, Singh & Panigrahi, 2017).

To mitigate the quality power-related problems during the integration of PV power, Voltage Source Converter (VSC) acts as a grid support converter allowing PV power to be effectively cared for in the network (Bhattacharjee & Roy, 2016; Naqvi, Kumar, & Singh, 2018). Moreover, VSC can be used to decrease harmonics and to improve the power factor via an ideal control. Literature shows only a few control strategies for VSC are proposed such as notch filter-based control direct sinusoidal tracer, and improved step channel-based control (Reddy & Ramasamy, 2018). Those strategies are acceptable only in the event of imbalance of the phase and particular rectification of frequency, not only for harmonics elimination. There are other strategies presented in the literature such as proportional integral based on the synchronous reference frame, Cauchy Schwarz inequality theory, Kalman filter, and notch-filtering approach (Shah & Singh, 2020; Sharma, Dahiya, & Nakka, 2019). Moreover, some research works present different adaptive procedures for calculation of amplitude and phase of a signal such as neuro-fuzzy inference, decentralized dynamic state estimation, least mean fourth, etc. Nevertheless, these approaches are acceptable under the condition of distorted grid voltage (Singh, Jain, Goel, Gogia, & Subramaniam, 2017; Tareen, Mekhilef, Seyedmahmoudian, & Horan, 2017, Malik, Fatema & Iqbal., 2021, Minai & Malik, 2020; Mahto et al., 2021). To estimate the phase angle, frequency, and amplitude, ANF dependent control system offers an instantaneous and exact response. However, the ANF method transient response is poor. Hence, it is not appropriate for the grid-integrated PV system. Besides, the Kalman filtering depends on a control method that provides exact outcomes in changeable frequency conditions. But, the selection of the covariance matrices is a major disadvantage. Furthermore, the WLSE depends control approach is robust and proficient in entire grid adverse conditions. Though, the complexity of the algorithm and the burden of computational are major disadvantages that limit the execution on low-cost microprocessors. For operating at an optimal level, it needs a robust control approach These aforementioned limitations are brought on by the mechanisms in place at the time that this research is conducted.

During offset of DC with the prevailing condition of harmonics, this approach cannot be able to control VSC of PV coordinated framework. To inject a balanced sinusoidal waveform of current into the grid, harmonic distortion or voltage irregularity conditions at the Point of Common Coupling (PCC), the Fundamental Component (FC) of PCC voltages should be eliminated (Chishti, Murshid, & Singh, 2019). Researchers have discussed many conventional control algorithms for single- and three-phase distribution grids. Synchronous reference frame theory (SRFT), and instantaneous reactive power theory (IRPT) (Puhan, Dash, Ray, Panda, & Pothauri, 2022). Several analysts have proposed a dual SOGI-based filter (Du-SOGI), a bandpass filter based on second order generalized integrator (SOGI), a cascade delay signal suppression method, an adaptive notch filter, a complex vector filter, and so on in this manner (Shah & Singh, 2018). But when the signal consists of DC component, or inter harmonics, then these techniques neglect to remove pure FC (Necaibia et al., 2018). To tackle this issue, several controllers such as Damped SOGI (DSOGI), integrator added SOGI, Second Order-SOGI (SO-SOGI), Modified SOGI (MSOGI), etc., are presented, which indeed forecast the DC segments, however, its presentation is undermined at high recurrence (Mensah, Yamoah, & Adaramola, 2019). In this manuscript, a new ANROA approach is proposed for a grid-tied SEC system for eliminating power quality problems like harmonics, voltage fluctuations and flickers, imbalance in loads, reactive power requirements, etc. The proposed approach decreases power quality problems with better accuracy. The suggested method combines the characteristics of both ANFIS and ROA; as a result, it is referred to as an ANROA control algorithm. Moreover, goal of presented control procedure was to utilize this solar power to satisfy load demand. Subsequent to fulfilling the requirement of the load, the remainder of solar power was fed into network. Features of the presented control method were approved via experimentation on created model. Through testing, unfavorable grid conditions, variety of solar-powered isolation, and load imbalance conditions were assumed to fulfill the intention of created control method.

Following are the primary contributions of the work that has been presented:

- A novel ANROA approach for a grid-tied SECS is proposed to avoid power quality problems such as harmonics, voltage deviation, load unbalance, reactive power management, etc.
- The ANFIS technique was incorporated to extract the maximum power from the solar PV system.
- The computational time of the ANROA technique is compared with P&O, PSO, LLMS, and ANF-based controllers, finding that the proposed ANROA technique reduces the power quality problem with less computational time.
- A statistical study of proposed and existing methodologies is performed finding that Total harmonic distortion (THD) for grid and load currents is reduced by 1.26 % using the proposed method.
- The validation of the proposed AFOGI-FLL controller is simulated in MATLAB/SIMULINK environment and FPGA-based experimental results are presented.

The rest of the paper is presented as follows: Sections 2 describes the system configuration; Section 3 presents the control strategy; Sections 4 depicts the proposed ROA for VSC control; Section 5 shows and discusses the results obtained; Section 6 presents the validation of the proposed AFOGI-FLL controller and shows the FPGA-based experimental results; Finally, Section 7 collects the conclusion of this work.

## 2. System Configuration

A typical standalone solar photovoltaic (SPV) generation system configuration connected to a two-stage grid is shown in Fig. 1. This configuration includes a DC-DC boost converter, SPV panel, ripple filter, boost converter, and three legs VSC. The SPV energy conversion system of the two-stage grid-connected is employed in this work. In the initial phase, solar PV modules are associated with the appropriate combination of series and parallel to obtain the preferred current and voltage rating of panels. A photovoltaic array is linked to the boost converter input for performing the Maximal Power Point Track (MPPT) function using the ANFIS approach and improving the array voltage to 700V when supplying power to the DC link. This procedure forces the solar PV array to reach the maximal power point at every condition of solar irradiance (Mao et al., 2020). In the second stage, the VSC is controlled through the ROA-based control technique. To interconnect the DC link by AC supply system, a three-leg VSC is used, which converts

DC power into appropriate AC voltages, load balance, reactive power compensation, and eliminates harmonic currents.

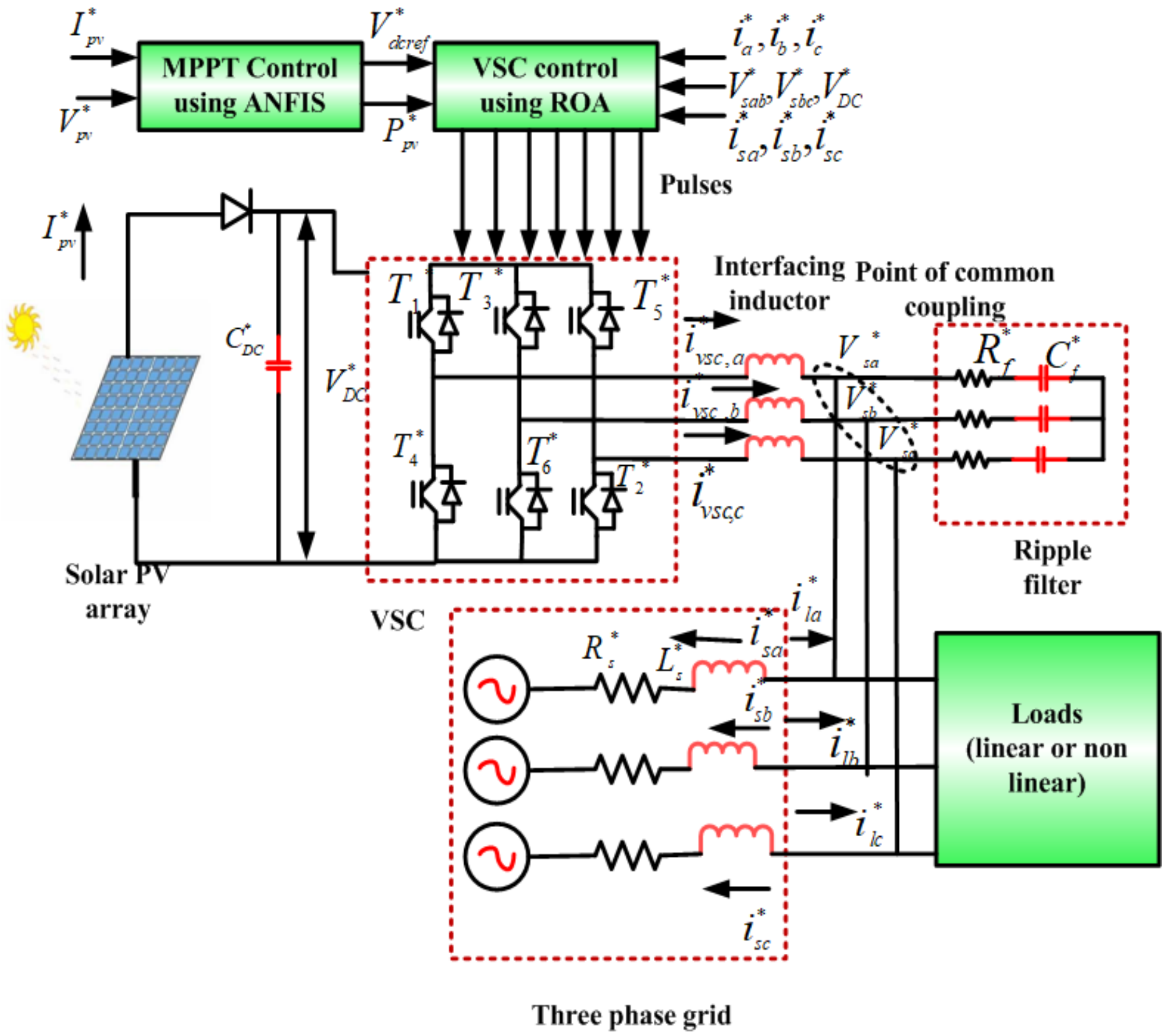

**Fig. 1.** System configuration of Grid-connected PV system.

Inductors of the VSC filter are utilized to interconnect the grid and VSC decreasing the ripple currents. Furthermore, a three-phase ripple filter is a shunt associated with PCC. The Ripple filter consumes the high-frequency switching noise produced via VSC. VSC switching pulses are generated using the VSC switching algorithm, which provides several functions such as power factor corrections, load balancing, reactive power management, and increased power utilization in the distribution network (Xin et al., 2016).

## 3. Control Strategy

A three-phase grid integrated solar PV system's control technique occurs in two-stage. Initially, the control strategy handles by a boost converter and then the control method handles by VSC, as illustrated in Fig. 2. The major scope of the proposed controller is producing the switching pulses of VSC to serve MPPT, and transferring the power from the PV field to the distribution system when preserving UPF operation. While the power absorbed through loads is greater than power extracted as solar PV string, the grid presents a positive resistance. The proposed control algorithm can be divided into two sections that name the MPPT together with control of the grid interface as the VSC part. MPPT algorithm converts the reference voltage of DC link and VSC interconnected control to the grid, which guarantees the performance of PV field in reference voltage of DC link together with power quality features (Mohamed, Jeyanthy, Devaraj, Shwehdi, & Aldalbahi, 2019). The DC link voltage ($V_{dc}^{*}$), photovoltaic current ($I_{pv}^{*}$), photovoltaic voltage $V_{pv}^{*}$, source/grid voltage $V_g^{*}$ and grid current $I_{pv}^{*}$, are sensed by the controlling system. The MPPT algorithm controls the boost converter. Therefore, $V_{dcref}^{*}$ is created based on the PCC voltage by equation (1). For improved dynamic performance and lower oscillations, the optimal step size is essentially required. As a result, Eq. 2 shows how the DC-link reference voltage is perturbed and how to compute the next step size to use the MPPT technique to acquire the most power possible.

$$V_{dcref}^{*} = \gamma\sqrt{2} \times V_{PA}^{*} \quad (1)$$

where the PCC voltage amplitude estimation and the value of $\gamma$ are greater than equal to one. The reference voltage is created depending on the conditions given by equation (2).

$$V_{dcref}^{*}(i) = \begin{cases} V_{dcref}^{*}(i-1) if \left( dV_{pv}^{*} \neq 0 \& \left( {dI_{pv}^{*}}/{dV_{pv}^{*}} \right) = -\left( {I_{pv}^{*}}/{V_{pv}^{*}} \right) \right) \\ V_{dcref}^{*}(i-1) + step if \left( dV_{pv}^{*} \neq 0 \& \left( {dI_{pv}^{*}}/{dV_{pv}^{*}} \right) > -\left( {I_{pv}^{*}}/{V_{pv}^{*}} \right) \right) \\ V_{dcref}^{*}(i-1) - step if \left( dV_{pv}^{*} \neq 0 \& \left( {dI_{pv}^{*}}/{dV_{pv}^{*}} \right) < -\left( {I_{pv}^{*}}/{V_{pv}^{*}} \right) \right) \\ V_{dcref}^{*}(i-1) - step if \left( dV_{pv}^{*} = 0 \& \left( dI_{pv}^{*} < 0 \right) \right) \\ V_{dcref}^{*}(i-1) + step if \left( dV_{pv}^{*} = 0 \& \left( dI_{pv}^{*} > 0 \right) \right) \\ V_{dcref}^{*}(i-1) if \left( dV_{pv}^{*} = 0 \& \left( dI_{pv}^{*} = 0 \right) \right) \end{cases} \quad (2)$$

Here, $V_{dcref}^{*}(i)$ is used to estimate the appropriate boost converter duty cycle. This duty cycle generates the pulse for the boost converters depending on the ANFIS approach. Reference grid current amplitude is given by equation (3).

$$I_{gp}^{*} = (I_l^{*} - I_f^{*}) \quad (3)$$

where $I_l^{*}$ and $I_f^{*}$ implied grid current loss component and feed-forward component respectively which are denoted by equations (4) and (5).

$$I_l^{*}(j) = I_l^{*}(j-1) + k_p^{*}((e_e^{*}(j) - e_e^{*}(j-1)) + k_l^{*} e_e^{*}(j) \quad (4)$$

$$I_f^{*} = \frac{2P_{pv}^{*}}{V_{PA}^{*}} \quad (5)$$

where $e_e^{*}$ is the error of the voltage which is varied through the PI-based controller and to produce grid current loss component, $P_{pv}^{*}$ is the PV power, and the system's dynamic performance can be enhanced depending on $k_p^{*}$and $k_l^{*}$ value. The grid reference current is calibrated by equation (6).

$$I_{gref}^{*} = I_{gp}^{*} \times u_p^{*} \quad (6)$$

The variance of $I_g^{*}$ is using the ROA approach, fed to the hysteresis controller to produce VSC switching pulses.

### 3.1. Proposed ANFIS for MPP Tracking

ANFIS is an artificial intelligence method used for different approaches in literature such as fault detection (Abbas & Zhang, 2021) and systems control (Weldcherkos, Salau, & Ashagrie, 2021). The advantage of this methodology is given by the neural networks as well as fuzzy systems. ANFIS method consists of a neural network with six layers (Moghadam, Izadbakhsh, Yosefvand, & Shabanlou, 2019), whose structure is shown in Fig. 3. Every layer has some nodes at node function: (a) process of Fuzzification; (b) Fuzzy rules; (c) Membership of standardization function; (d) Fuzzy rules of resultant piece; (e) Network exit. For deciding the parameters of the fuzzy inference system, the ANFIS structure of the Sugeno hybrid training algorithm is employed. The methodology uses least-squares methods as well as the BP gradient descent algorithm to create and manage parameter units of membership functions in a fuzzy inference system (Moghadam et al., 2019). In this paper, the ANFIS approach is selected for extracting the maximal power as SECS by frequently varying load conditions to control the boost converter. PV current $I_{pv}^{*}$and PV voltage $V_{pv}^{*}$ are considered ANFIS input systems. The condition for tracking MPP are given by equations (8) and (9).

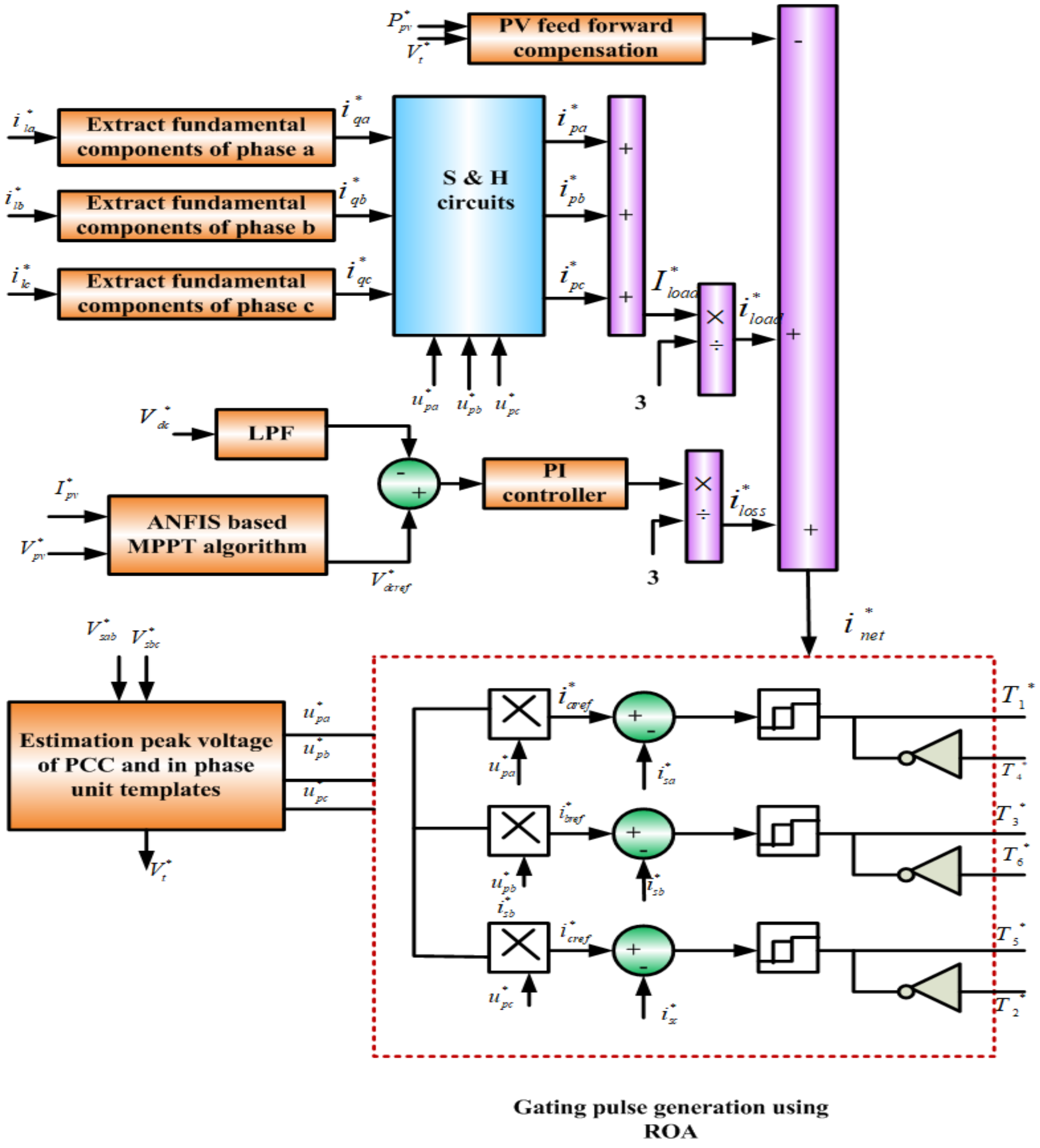

**Fig. 2.** Schematic diagram of control strategy.

If $\frac{dI_{pv}^*}{dV_{pv}^*} > -\frac{I_{pv}^*}{V_{pv}^*} \Rightarrow D^* = D_{old}^* - \Delta D^*$ (7)

If $\frac{dI_{pv}^*}{dV_{pv}^*} = -\frac{I_{pv}^*}{V_{pv}^*} \Rightarrow D^* = D_{old}^*$ (8)

If $\frac{dI_{pv}^*}{dV_{pv}^*} < -\frac{I_{pv}^*}{V_{pv}^*} \Rightarrow D^* = D_{old}^* + \Delta D^*$ (9)

where $D_{old}^*$ and $\Delta D^*$ are the duty ratio of earlier boost converter, and the step change in the duty ratio respectively.

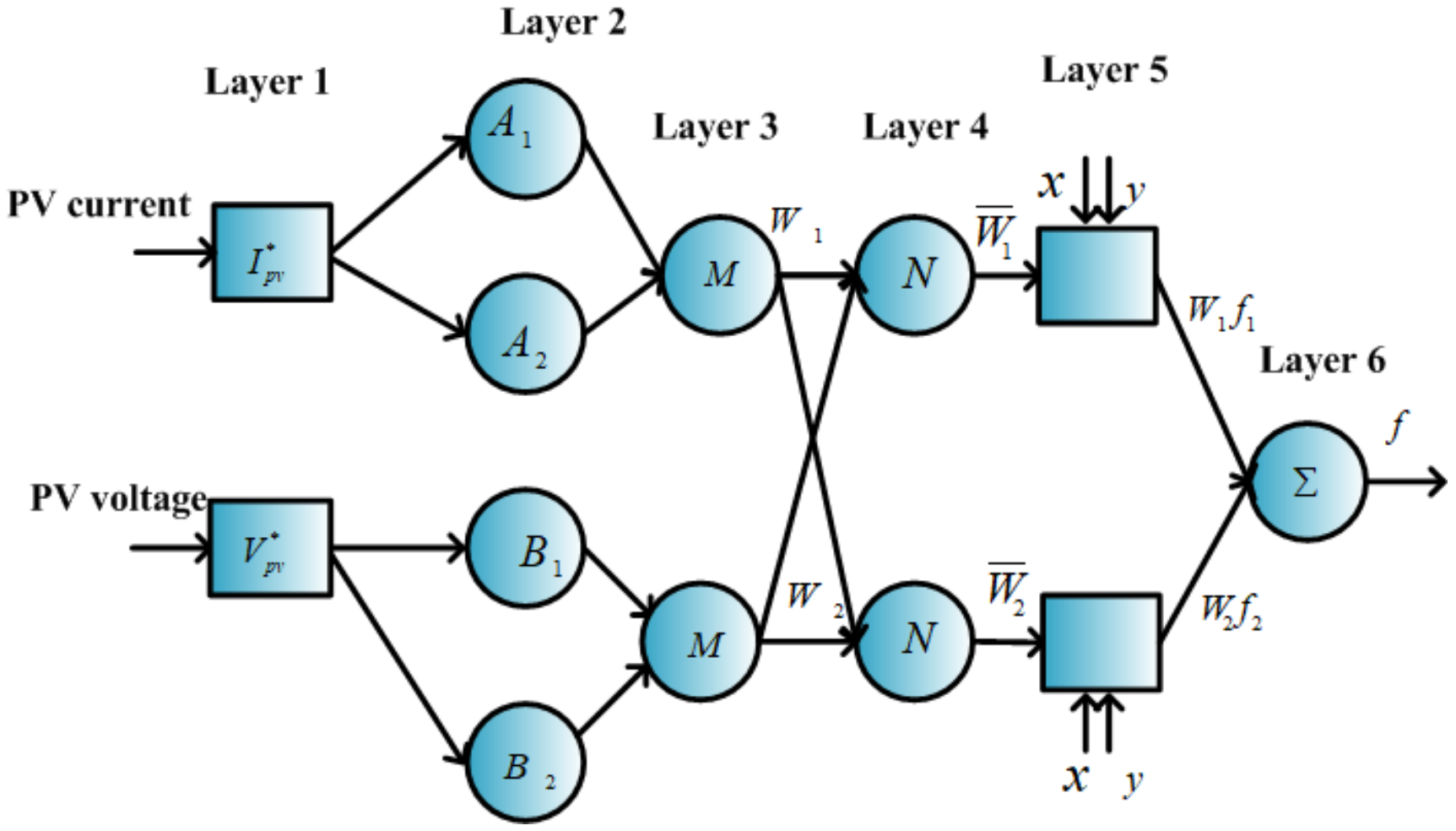

**Fig. 3.** Structure of the ANFIS.

The first step is to configure a fuzzy rule, determining the input variables as well as fuzzification. Each node represents a node from adaptive along functions of the node explained as pursued in this layer. The outcome function of layer 2 is estimated using equation (10),

$O_j^1 = \mu^{A_j}(X),\ j = 1, 2$ (10)

where input of $j^{th}$ node implies $X$ as well as along the node is $A_i$ a fuzzy set is interrelated. Therefore, the degree of membership in fuzzy sets $(A_1, A_2)$ is denoted as $O_j^1$, and the following input $X$ is fulfilled by quantifier $A$. The function of membership for fuzzifying based on the Bell function is denoted in equation (11).

$\mu^{A_j}(x_1) = \frac{1}{1+\left(\frac{x-d_j}{a_j}\right)}$ (11)

Here, $d_j$ and $a_j$ implied the Bell function parameters. Layer 3 is resoluted using “M” as a circle node. Moreover, the firing strength of every rule is estimated at this layer. The output layer can be calculated by equation (12),

$O_j^2 = W_j = \mu^{A_j}(X) \cdot \mu^{B_j}(Y), \quad j = 1, 2$ (12)

Each T-norm operator performs a fuzzy operation in this layer. Every node is denoted “N” for layer 4. Output node implies $O_j^3\ or\ \overline{W_j}$, and it is calculated by equation (13),

$O_j^3 = \overline{W_j} = \frac{W_j}{W_1+W_2},\ j = 1, 2$ (13)

For layer 5, every node is an adaptive square node with the function of the node, which is given by equation (14).

$$O_j^4 = \overline{W_j} \cdot f_j = \overline{W_j} \cdot (p_j x + q_j y + r_j), \quad j = 1, 2 \tag{14}$$

Here, $f_j = p_j x + q_j y + r_j$ of the $j^{th}$ node. A single node estimates the total number of the entire received signals for layer 6 to produce the overall outcome. ANFIS output is similar to the unique node, and it is modeled by equation (15).

$$O_j^5 = f = \sum_j W_j f_j = \frac{\sum_j W_j \cdot f_j}{\sum_j W_j} \tag{15}$$

ANFIS structure is shown in Fig. 4. In this research, the ROA technique is used to control the VSC in the ANFIS output network.

## 4. Proposed ROA for VSC Control

The Rain Optimization Algorithm (ROA) is a novel algorithm that is triggered by raindrops that move towards minimal spots after contact. Also, ROA may locate the local minimal concurrently, as well as it may be utilized confidently at optimization issues. Based on this issue, a few points on response space may be chosen arbitrarily when raindrops fall to the ground. The radius of each raindrop is its most important characteristic. The radius of each raindrop may be decreased with time and may be maximized as one raindrop connects with other raindrops. While the primary population of responses is generated, the radius of every drop may be randomized into a suitable range. With each repetition, each drop verifies its neighborhood based on its size. For individual drops that are not yet linked to any other drops, just verify the final boundary of the place covered. The solved problem has a space of n dimensions, and each drop has n variables. The lowest and maximum limits of variable one will then be examined in the first phase, as these constraints can be established by the drop radius. Over the next phase, two endpoints of variable two will be evaluated, and so on up to the last variable is tested. At this phase, the cost of the initial drop would be upgraded by moving down. This action will be done for entire drops, after that cost and position of entire drops would be allocated (Moazzeni & Khamehchi, 2020). Moreover, the procedural steps of the ROA algorithm are shown in Fig. 4. The ROA methodology is being used to control the grid-connected VSC in this paper. This step-by-step process involved in making switching pulses are explained in the following subsections.

### 4.1. ROA Step-by-Step Process for VSC Control

Every solution to the problem can be modeled through a raindrop in this approach. PCC voltage $V_t^*$, grid current $I_g^*$, load current $I_L^*$, and PV current $I_{pv}^*$ PV voltage $V_{pv}^*$ are taken as the input parameters for the ROA technique. The input parameters are created at random after the startup process. The fitness values are calculated for randomly generated parameters. If the value of fitness is less than zero, then a switching pulse is generated, otherwise, the system does not generate the switching pulses. The generated switching pulses control the VSC. The reference current is evaluated in the proposed control algorithm, utilized to generate VSC switching pulses. The detailed step process is explained as follows,

**Step 1: Initialization**

First of all, PCC voltage $V_t^*$, grid current $I_g^*$, load current $I_L^*$, PV current $I_{pv}^*$and PV voltage $V_{pv}^*$ are initialized.

**Step 2: Random Generation**

Following the setup phase, the input parameters are randomly generated. Depending on the fitness function, the highest fitness values are picked in this stage. The three-phase voltages are evaluated using sensing line voltages $V_{sab}^*$ and $V_{sbc}^*$ by equations (16)-(18).

$$V_{sa}^* = \frac{2V_{sab}^* + V_{sbc}^*}{3} \quad (16)$$

$$V_{sb}^* = \frac{-V_{sab}^* + V_{sbc}^*}{3} \quad (17)$$

$$V_{sc}^* = \frac{-V_{sab}^* - 2V_{sbc}^*}{3} \quad (18)$$

The PCC voltage average amplitude is estimated using equation (19).

$$V_t^* = \frac{\sqrt{2\left(V_{sa}^{*2} + V_{sb}^{*2} + V_{sc}^{*2}\right)}}{3} \quad (19)$$

In phase unit templates are attained using equations (20)-(22).

$$u_{pa}^* = \frac{V_{sa}^*}{V_t^*} \quad (20)$$

$$u_{pb}^* = \frac{V_{sb}^*}{V_t^*} \quad (21)$$

$$u_{pc}^* = \frac{V_{sc}^*}{V_t^*} \quad (22)$$

The PV feed-forward component is calculated by equation (23)

$$I_{pvf}^* = \frac{2P_{pv}^*}{3V_t^*} \quad (23)$$

**Step 3: Fitness Function**

From the initialized values, the solution random number is generated. The solution's fitness function is assessed, and it is described by equation (24):

$$F_i = \begin{cases} \frac{1}{1+F_i} & if\ \ F_i < 0 \\ 1 + bcs(F_i) & if\ \ F_i < 0 \end{cases} \quad (24)$$

If $F_i < 0$, then the switching pulse is generated for controlling the VSC, and, if $F_i > 0$, the mean switching pulse cannot be generated.

**Step 4: Generation of Switching Pulses**

Two droplets among radius $r_1$ and $r_2$ are closed with everyone, which consists of the general area by everyone; they may outline switching signal based on equation (25).

$$R^* = \left(r_1^{n^*} + r_2^{n^*}\right)^{1/n^*} \quad (25)$$

Here, $n^*$ implies a number of variables on every droplet. Error among sensed grid current, as well as their reference value, are employed for generating switching pulses. Sensed grid currents are represented by equations (26)-(28)

$$i_{sa}^* = i_{pa}^* + i_{qa}^* \quad (26)$$

$$i_{sb}^* = i_{pb}^* + i_{qb}^* \quad (27)$$

$$i_{sc}^* = i_{pc}^* + i_{qc}^* \quad (28)$$

**Step 5: Control the VSC**

Error among reference grid $I_{gref}^*$ and sensed grid $i_{sa}^*$, $i_{sb}^*$ and $i_{sc}^*$ are passed via the hysteresis current control, which produces gating pulses of switching of the converter. If a drop-through radius $r_1$ not move in terms of soil properties, then VSC is controlled based on equation (29).

$$R^* = \left(r_1^{n^*}\beta\right)^{1/n^*} \tag{29}$$

The process continues until the best solution is found. Fig. 4 depicts the flow chart for the proposed ANROA approach.

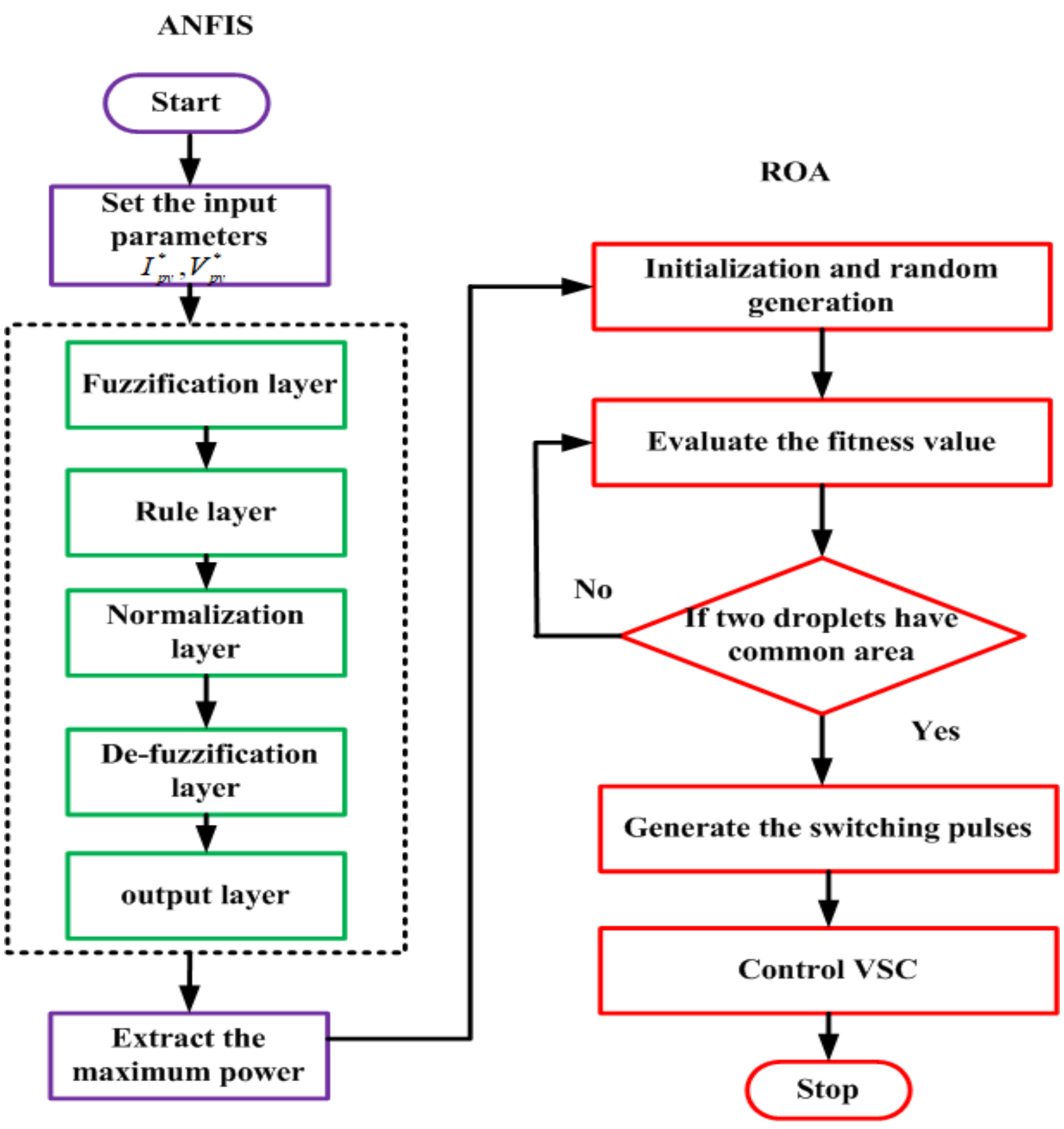

**Fig. 4.** Flow chart of the proposed ANROA approach.

## 5. Results and discussion

In this manuscript, a novel ANROA method is proposed for 3-phase PV (photovoltaic) system connected to the AC grid. The proposed method refers fusion of the ANFIS and ROA method and, hence, it is called as ANROA approach. Primarily, maximal power point is tracked by ANFIS technique. After that, the ROA approach is utilized to generate the getting pulses to VSC. Moreover, ROA approach extracts the basic component of grid voltage, even, while grid voltage is categorized via over-voltage, variations of the frequency, under-voltage, serious harmonic distortion, direct current, and so on, in a broad range. The major scope of this system is removing power quality problems, like voltage fluctuations, harmonics together with flickers, unbalance in loads, and reactive power requirement. Besides, the proposed system is performed at zero voltage regulation and unity power factor modes. The MATLAB / Simulink environment is used to perform the proposed model. Table 1 lists the system parameters that were employed in the simulation.

**Table 1.** Simulation parameters.

| Parameters | Ratings |
|---|---|
| PV array power | 32.5 kW |
| PV module current | 7.61 A |
| PV module voltage | 26.3 V |
| Grid voltage | 415 V |
| Ripple filter | 5Ω, 10 μF |
| $k_p^*$ | 4.8 |
| $k_l^*$ | 1.0 |
| DC link voltage | 710 V |
| Interfacing inductor | 2.5 mH |

In this paper, the PV array current and voltage values are taken as 7.61 A and 26.3 V. The value of PV array power is taken as 32.5 kW. The grid voltage is 415 V, and the interface inductor has a rating of 2.5 mH. Furthermore, $k_p^*$ and $k_l^*$ value is taken as 4.8 and 1. The performance of the proposed ANROA technique is investigated in balanced as well as in unbalanced non-linear load and for PV partial shading conditions. To analyze the effectiveness of the proposed approach, the results are compared to some of the most recent approaches, such as P&O, PSO (Particle swarm optimization), LLMS, and ANF.

**Case 1: Balanced Nonlinear Load condition**

In this case, 32.5 kW solar PV system interfaced non-linear load is associated with the PCC (Point at common coupling) terminals. The VSC current contains the harmonics components which are identical and opposite to the load current for managing the primary frequency in the grid current. Test outcomes for the operation of the SPV system interconnected to a two-stage three-phase grid in non-linear load conditions are demonstrated in Fig. 5 where photovoltaic power (Ppv), Photovoltaic current (Ipv), DC link voltage (Vdc), feed-forward current (Iff), grid power (Pg), and reactive power of grid (Qg) are represented. Reactive power requirement for non-linear load obtained via the PV solar array is linked to VSC in PCC. Furthermore, reactive power flow to the grid is zero, illustrating that the distribution network operated at a unity power factor, also it is noticed that Vdc is maintained constant at 1000 W/$m^2$ irradiance level is 700 V. Fig. 6 shows grid voltages (Vgabc), grid currents (Igabc), constant irradiance (Ir), load currents (iLa, iLb, and iLc), and currents in balanced non-linear load situations. It is shown in Fig. 6 that at constant solar irradiance grid voltages and grid currents are exactly balanced and sinusoidal. Therefore, reactive power is compensated with unity power factor operation under non-linear loading conditions.

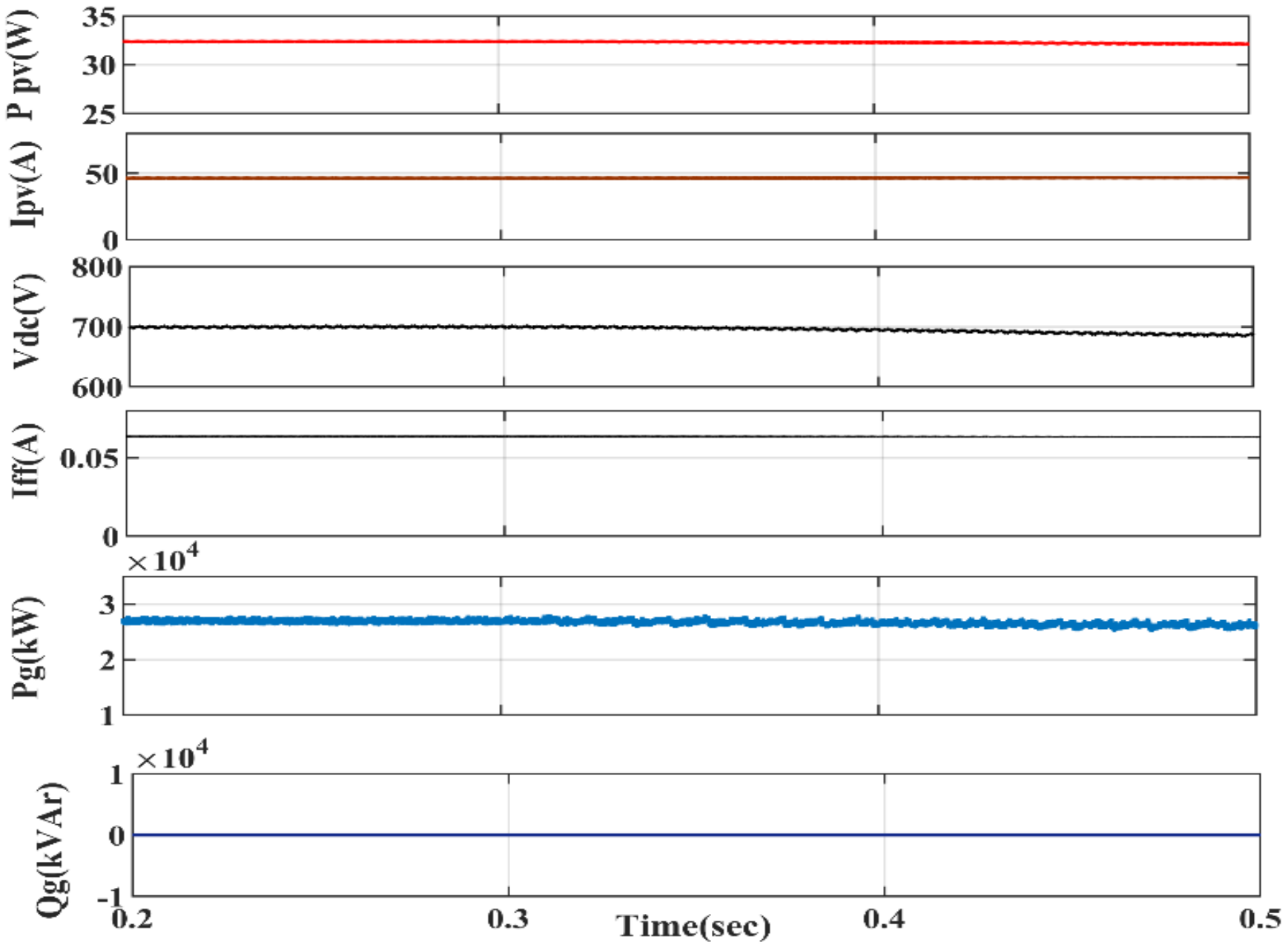

**Fig. 5.** PV power (Ppv), PV current (Ipv), DC link voltage (Vdc), Feed-forward current (Iff), grid active power (Pg) and reactive power (Qg), under balanced non-linear load condition.

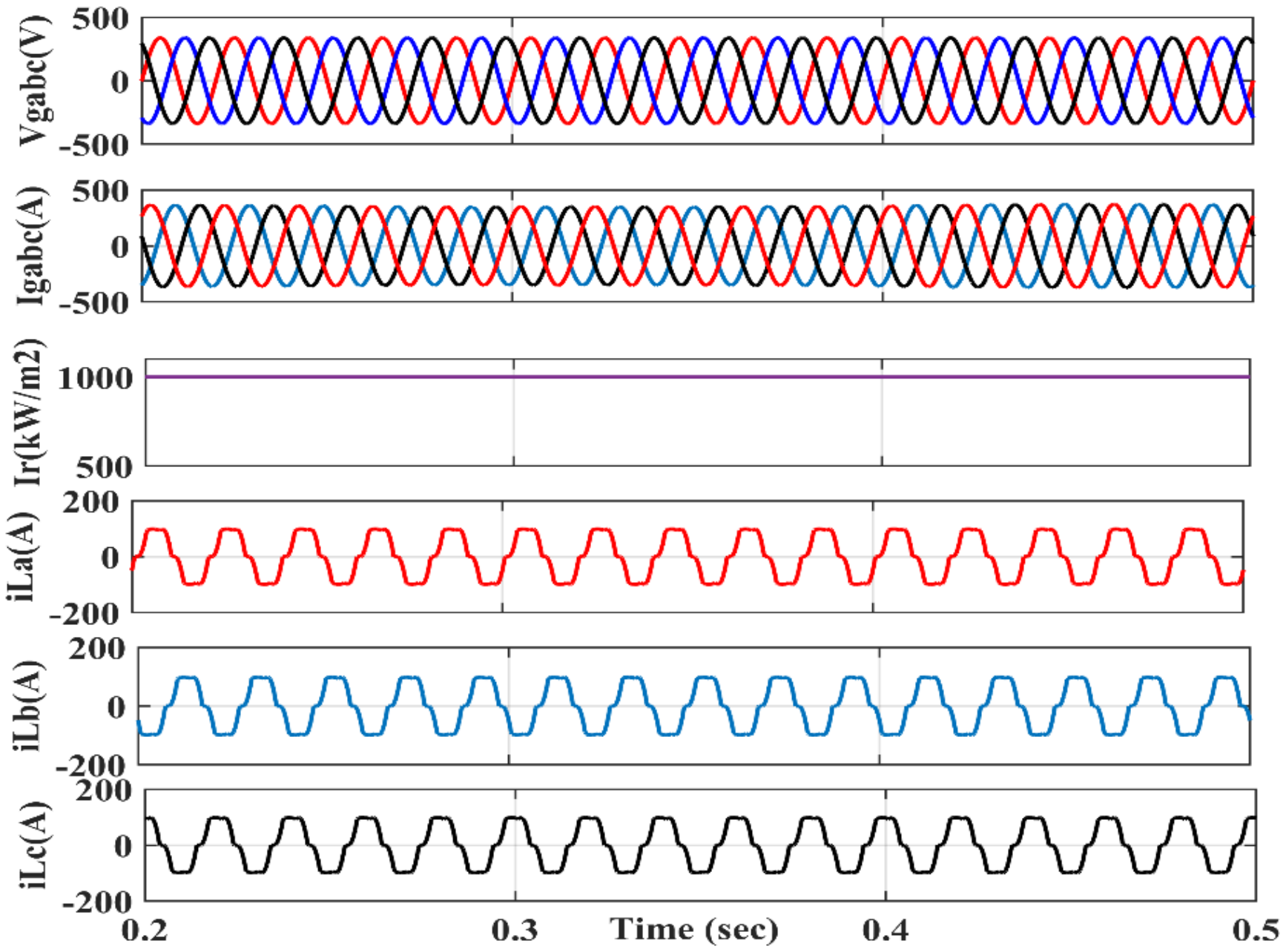

**Fig. 6.** Grid voltage (Vgabc), Grid Current (Igabc), constant irradiance (Ir) and load currents (iLa, iLb, iLc) under balanced non-linear load condition.

**Case 2: Unbalanced Nonlinear Load Condition**

To emulate a non-linear load, the diode bridge are considered by RL load on the DC side. The SPV (Solar photovoltaic system) performance under nonlinear load conditions is shown in Fig. 7. In this case, unbalancing occurs on the load side. One of the phases is removed between time t=0.3 sec to time t=0.4 sec. The grid voltage (Vgabc) and grid current (Igabc) are sinusoidal and balanced with constant irradiance at this time instant (Ir). Fig. 7 shows the waveforms of load currents (iLa, iLb and iLc), when phase 'a' is disconnected (iLa), the nature of the other two-phase are the same. Fig. 8 depicts the DC link voltage, solar PV power remains constant. Moreover, it is observed that the reactive power (Qg) flow on the grid is zero, and the system is operated at a unity power factor. Additionally, by removal of load does not affect the SPV power (Ppv), voltage, and current (Ipv). Moreover, in respect of load disconnection of one phase DC link voltage (Vdc) is maintained constant with very minute variation at time t=0.3 sec to t=0.4 sec. Also, it is noticed that the grid power (Pg) is decreased at that instant.

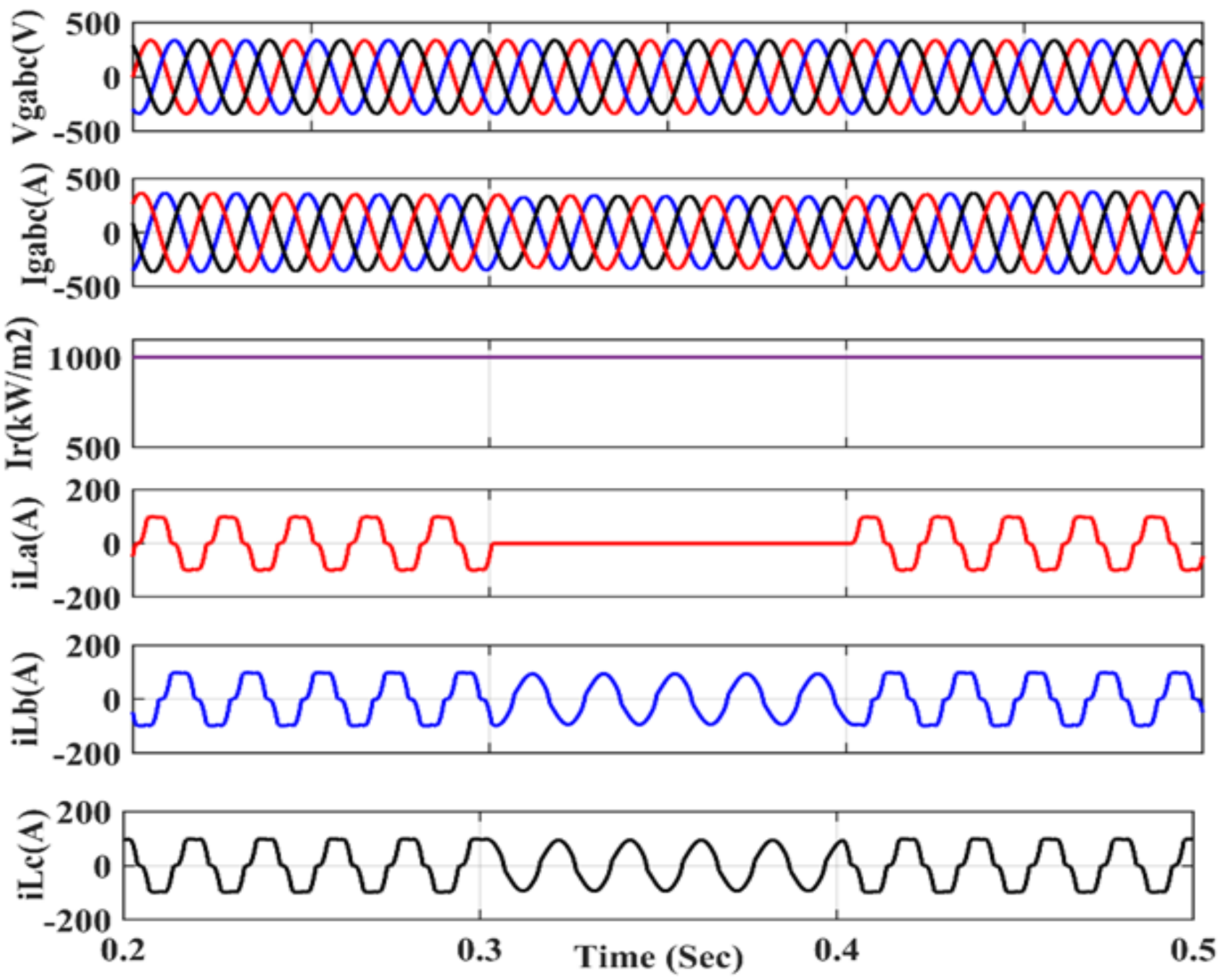

**Fig. 7.** Unbalanced non-linear load state, grid voltage, grid, load, inverter, and reference currents behavior.

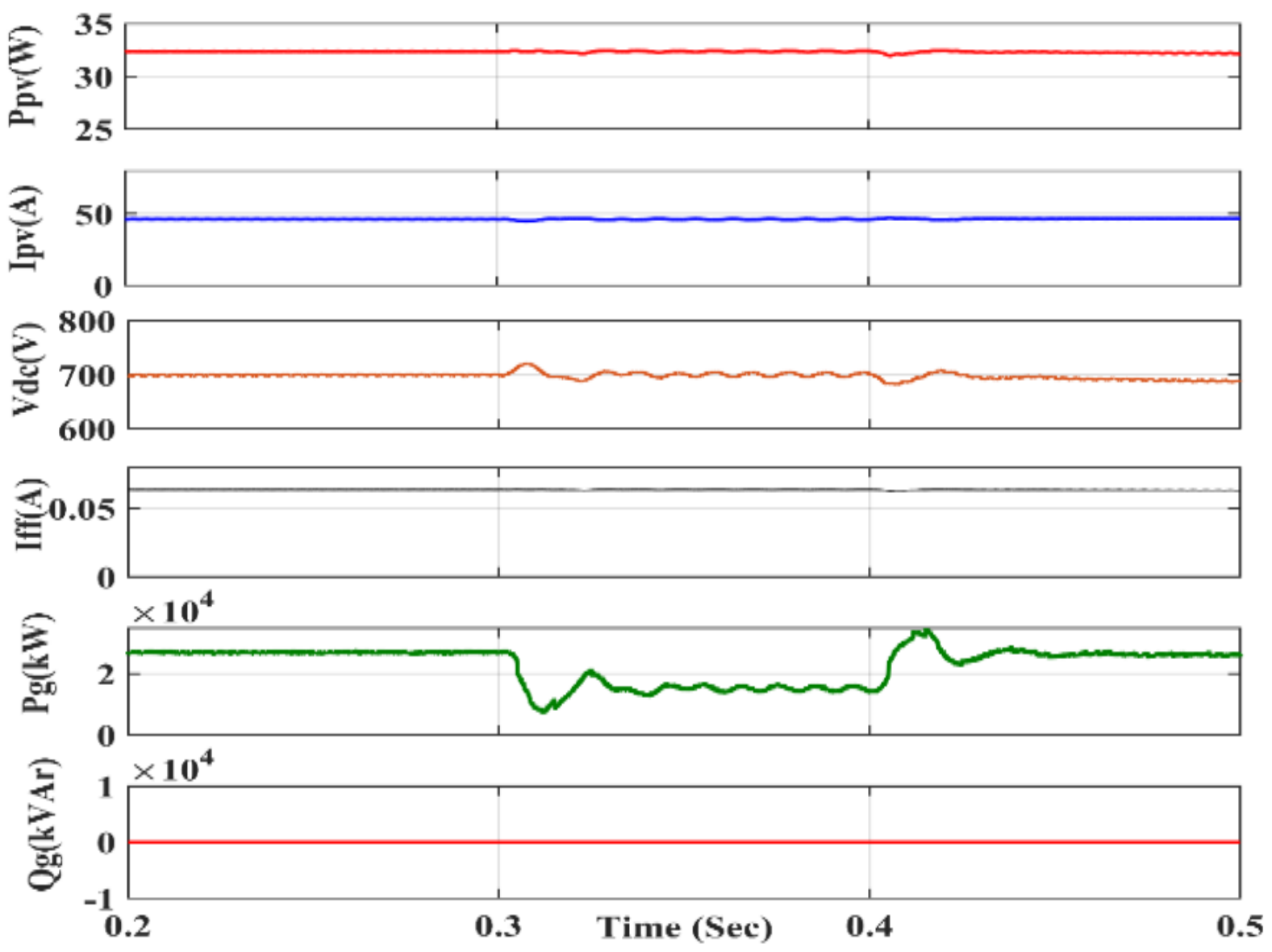

**Fig. 8.** PV current, load currents, DC link voltage, PV, and grid power in unbalanced non-linear load condition.

**Case 3: PV Partial Shading Condition**

The behavior of the system in PV partial shading conditions with non-linear load is displayed in Fig. 9. This figure depicts the waveforms of PV power (Ppv), photovoltaic current (Ipv), DC link voltage (Vdc), grid instantaneous power (Pg), and grid reactive power (Qg). It is shown that SPV system power is reduced after at t=0.3sec due to a change in irradiance. As a result of the irradiance variation, the AC grid feeds the remaining active power demand of the load, and there is only a minor burden of reactive power on the AC grid. Moreover, DC link voltage remains constant due to partial shading conditions. Moreover, due to partial shading, grid voltage and current are perfectly balanced and sinusoidal. Fig. 10 depicts the nature of grid voltages (Vgabc), currents (Igabc), compensating current for phase ‘a’ (Ica), and load current of phase ‘a’ (Ila). It is noticed that irrespective of partial shading conditions grid voltages and grid currents are remains balanced. Also, the system is operated at a unity power factor. In addition, Fig. 11 and Fig. 12 show the THD of grid current and load current, respectively. The grid's current THD is well within the IEEE standard.

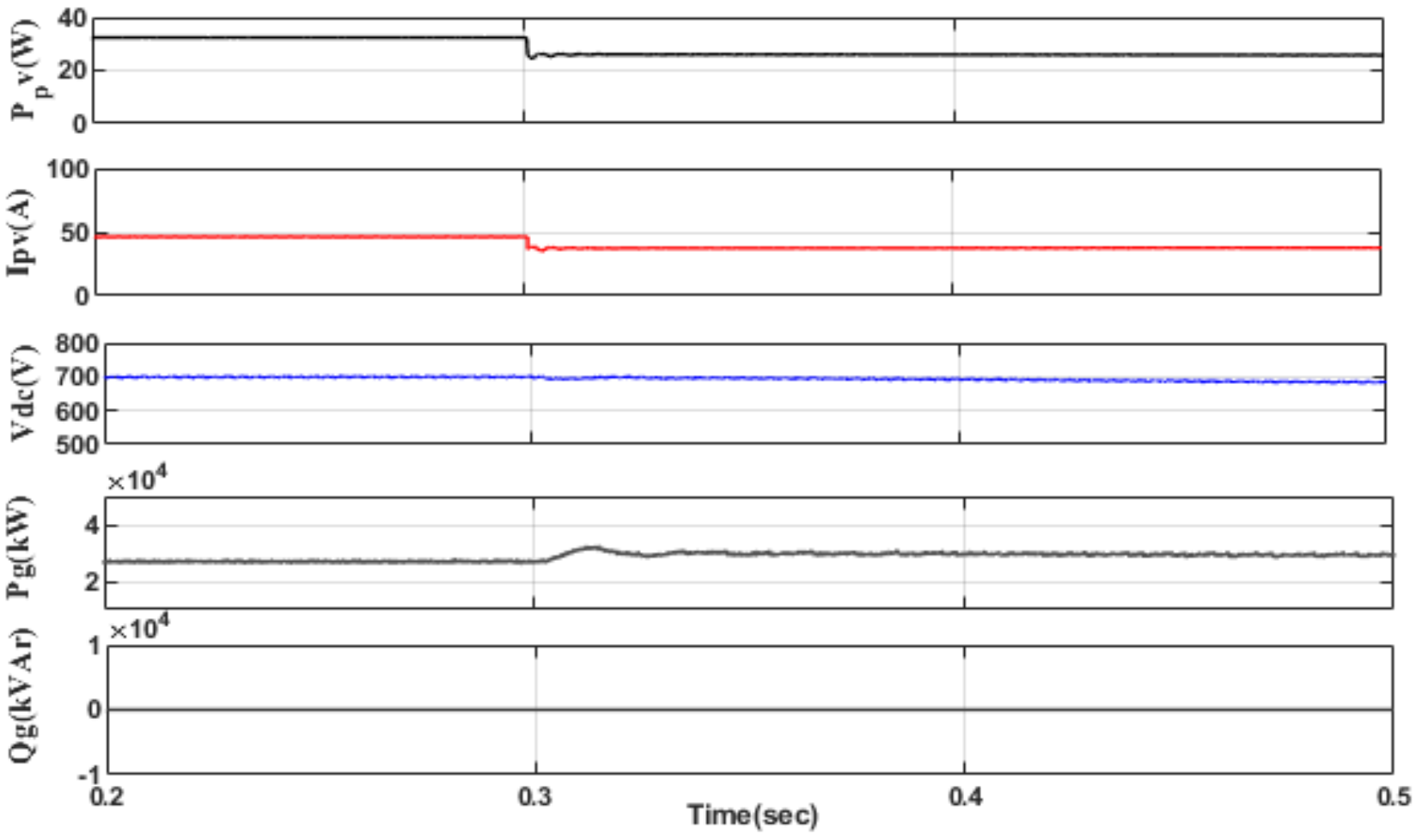

**Fig. 9.** PV current, load currents, DC link voltage, PV and grid power under balanced nonlinear Load condition.

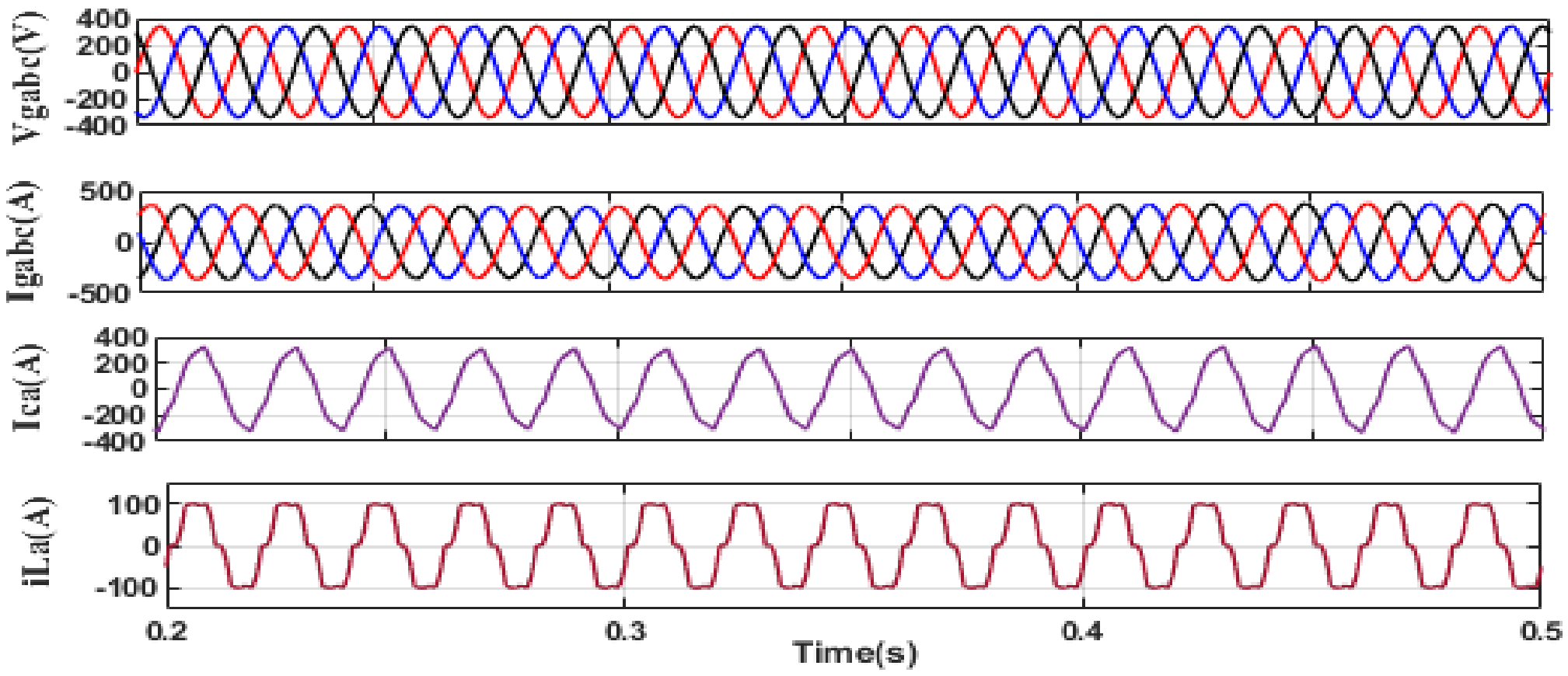

**Fig. 10.** Grid voltage, grid, load, inverter, and reference currents under balanced nonlinear Load condition.

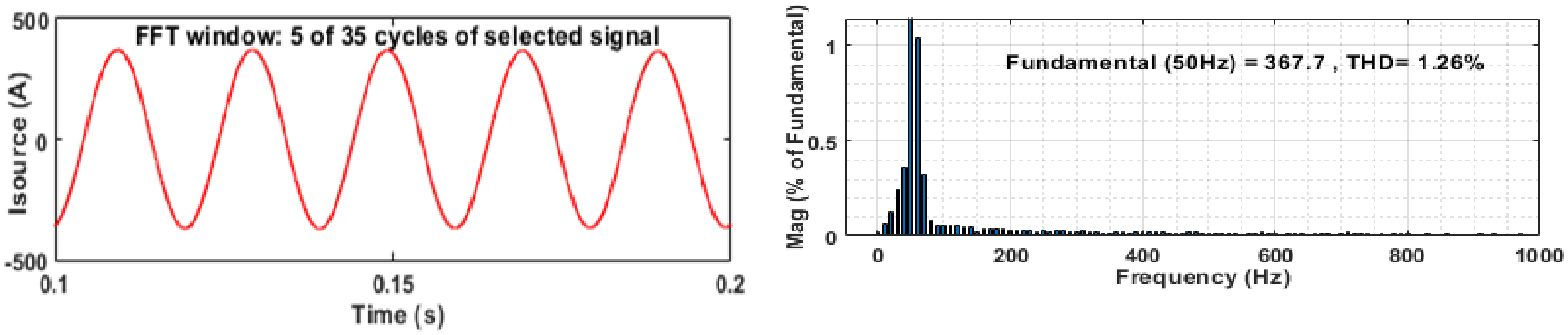

**Fig. 11.** Harmonic spectrum for grid current.

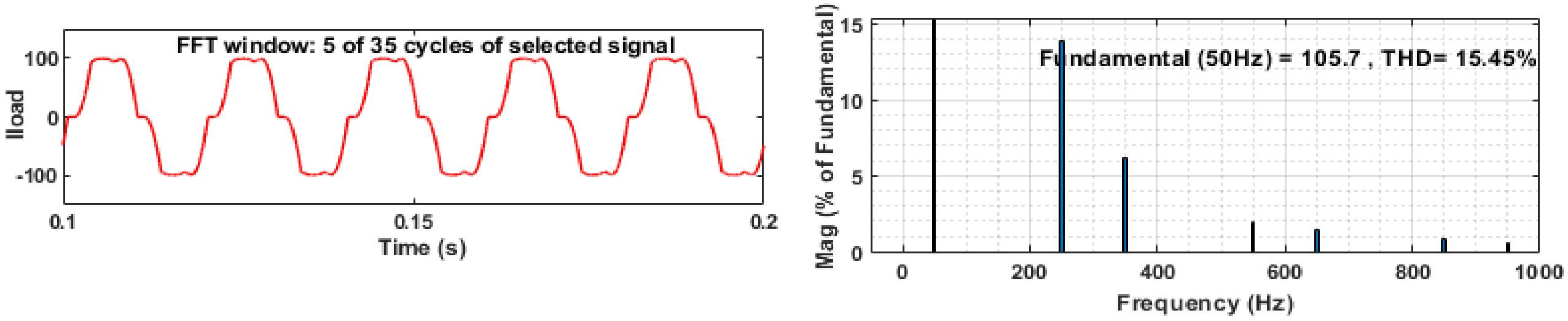

**Fig. 12.** Harmonic spectrum for load current.

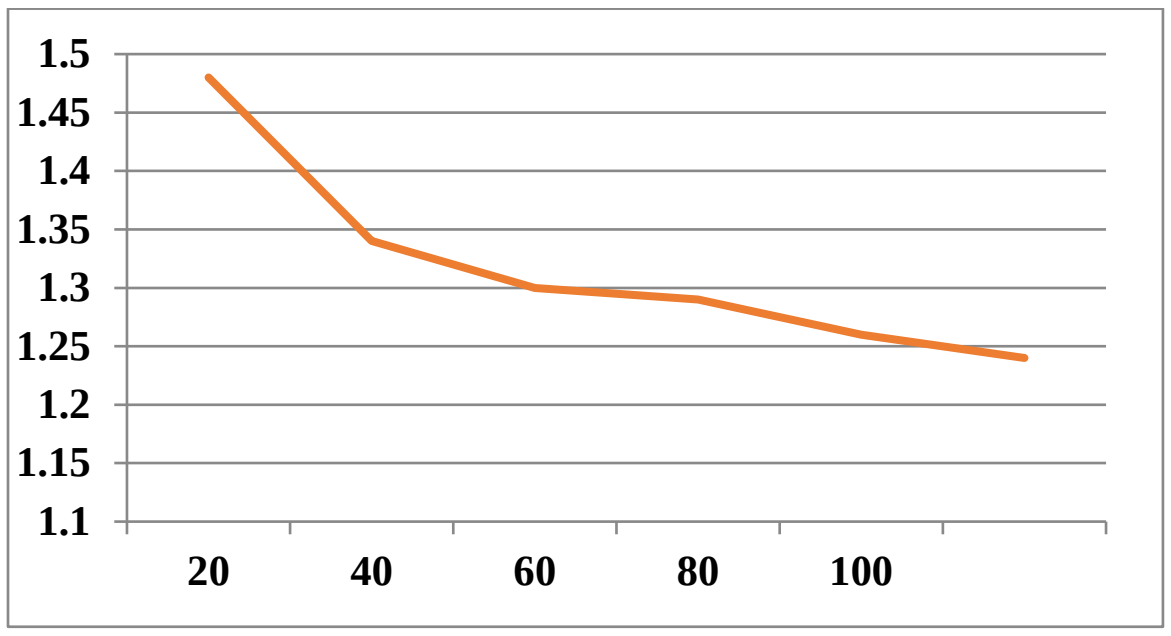

**Fig. 13.** Convergence curve for the proposed system.

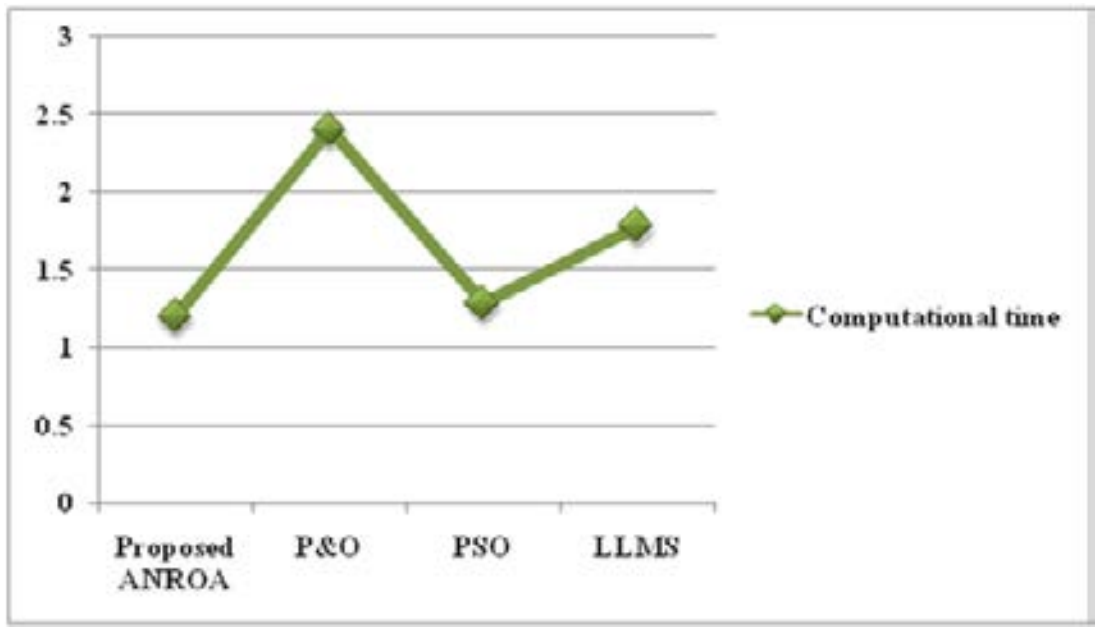

**Fig. 14.** Comparison of suggested and existing methodologies in terms of computational time.

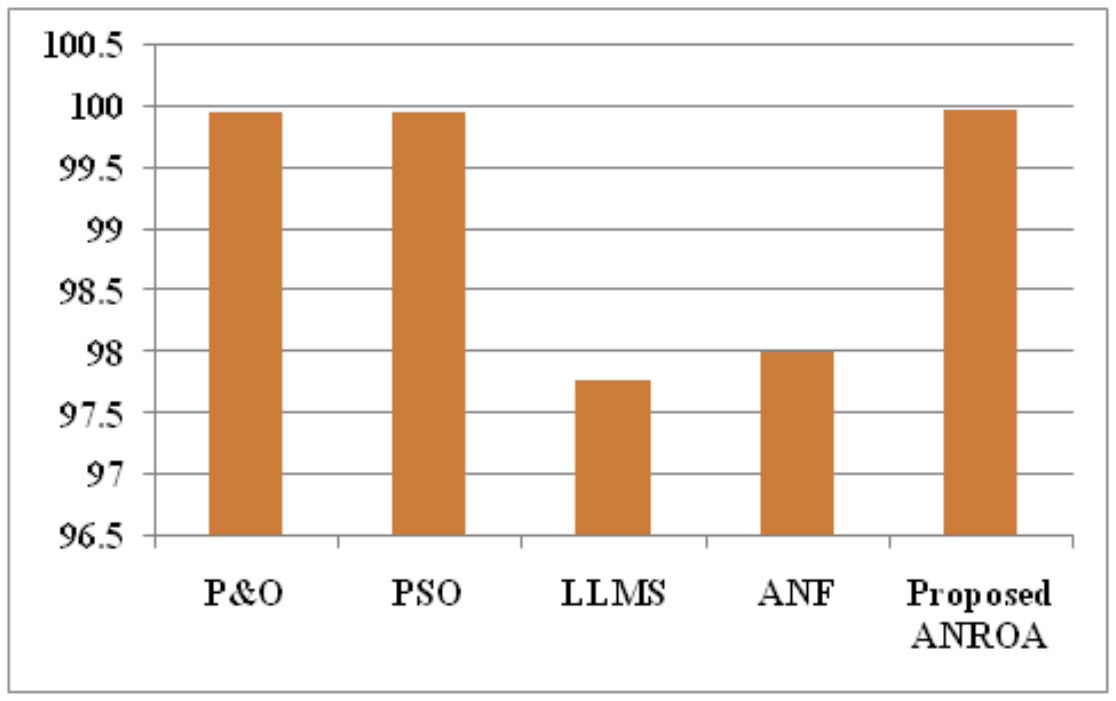

**Fig. 15.** Comparison of tracking proposed efficiency without partial shading.

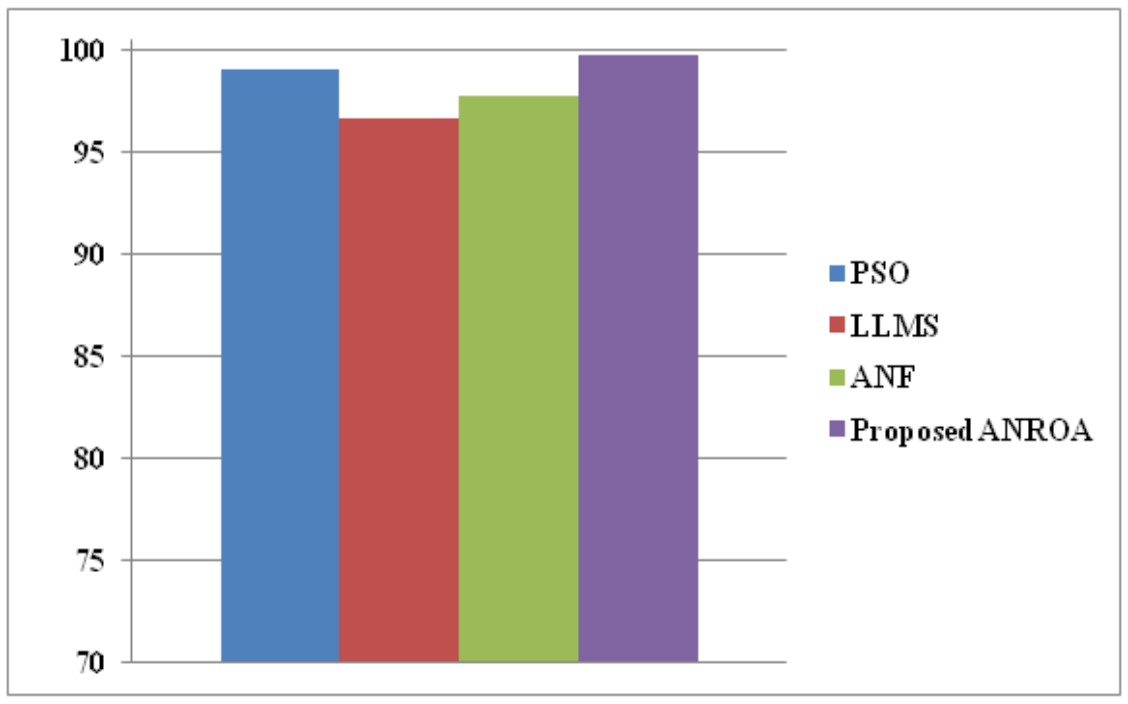

**Fig. 16.** Comparison of tracking proposed efficiency with shading.

**Table 2**. Measures of proposed and existing techniques based on statistics.

| Method | Mean | Median | S.D |
|---|---|---|---|
| Proposed ANROA | 1.2561 | 1.2036 | 0.0971 |
| P&O | 1.2890 | 1.2217 | 0.1019 |
| PSO | 1.3094 | 1.2345 | 0.1084 |
| LLMS | 1.2887 | 1.2452 | 0.1564 |
| ANF | 1.3170 | 1.3345 | 0.1214 |

**Table 3.** Performance analysis with partial and without partial shading.

| | Without partial shading | | | | | With partial shading | | | | |
|---|---|---|---|---|---|---|---|---|---|---|
| | P&O | PSO | LLMS | ANF | Proposed ANROA | P&O | PSO | LLMS | ANF | Proposed ANROA |
| Time to track MPP (s) | 15 | 3.56 | 34 | 11.86 | 0.355 | 15 | 4.78 | 35.780 | 13.768 | 0.378 |
| Power oscillation in steady state (%) | 0.020 | 0.009 | 0.060 | 0.078 | 0.018 | 0.060 | 0.008 | 0.980 | 0.60 | 0.016 |
| Tracking efficiency (%) | 99.96 | 99.96 | 97.78 | 98 | 99.98 | 71.93 | 99.05 | 96.67 | 97.78 | 99.801 |

**Table 4.** Comparison of suggested and existing approaches in terms of performance.

| **Method** | **Computational time (s)** | **Sampling time (µs)** | **Accuracy (%)** |
|---|---|---|---|
| Proposed ANROA | 1.202 | 20 | 99.96 |
| P&O | 2.40 | 50 | 98.56 |
| PSO | 1.28 | 50 | 99.76 |
| LLMS | 1.78 | 30 | 99.67 |
| ANF | 3.26 | 42 | 95.23 |

**Table 5.** THD comparison of proposed and existing approaches.

| **Method** | **THD** | |
|---|---|---|
| | **Grid current (%)** | **Load current (%)** |
| Proposed ANROA | 1.26 | 15.45 |
| P&O | 2.45 | 24.43 |
| PSO | 8.12 | 17.67 |
| LLMS | 1.63 | 24.43 |
| ANF | 4.26 | 13.34 |

The harmonics spectrum for grid current is shown in Fig. 11. Grid current THD is around 1.26%. The harmonics spectrum for load current is shown in Fig. 12. Load current THD is around 15.45 %. All load current lower order harmonics are produced via the SPV system and therefore grid current THD is lower when compared with load current. The convergence curve of the proposed ANROA method is shown in Fig. 13. A convergence value of 1.48 is achieved using a suggested Kernel PCA-ESMO technique. In addition, Fig. 14 shows a comparison of the suggested and existing methodologies in terms of computational time. With less computational time, the proposed ANROA techniques reduce the power quality problem. The computational time of the P&O, PSO, LLMS, and ANF is 2.4 sec, 1.28 sec, 1.78 sec, and 3.26 sec, respectively. The tracking proposed efficiency comparison without partial shading is shown in Fig. 16. A proposed system tracking efficiency is 99.98% without partial shading conditions. The tracking proposed efficiency comparison with partial shading is shown in Fig. 17. A proposed system tracking efficiency is 99.80 % with partial shading conditions. Statistical analysis of proposed and existing techniques is shown in Table 2. The proposed technique's mean value is 1.2561, the median is 1.2036, and the standard deviation is 0.0971. These values are the lowest, compared to the existing techniques. The performance analysis with partial and without partial shading is demonstrated in Table 3. Table 4 shows the performance comparison proposed by existing techniques. The computational time of the P&O, PSO, LLMS, and ANF is 2.4 sec, 1.28 sec, 1.78 sec, and 3.26 sec respectively. The proposed ANROA technique achieves a 1.202-sec computational time, which is low when compared to the conventional approaches. Furthermore, the accuracy of the proposed system is 99.96 %. Table 5 shows the THD comparison of planned and existing techniques. The proposed system attains 1.26 % THD for grid current and 15.45 % THD for the load current. As a result, it is concluded that the new strategy outperforms the existing approaches.

## 6. Experimental Results

The proposed controller is analyzed also through the hardware in the loop (HLL), FPGA based real-time simulator. The validation and evaluation of the proposed AFOGI-FLL controller setup are shown in Fig.17. The experimental set-up comprises of target simulator (OP RTS 5700), a Host simulator, UPS supply, Mixed-signal oscilloscope (MSO), CAN( Controller Area Network) bus and a BNC (Bayonet Neil-Concelman) cable.

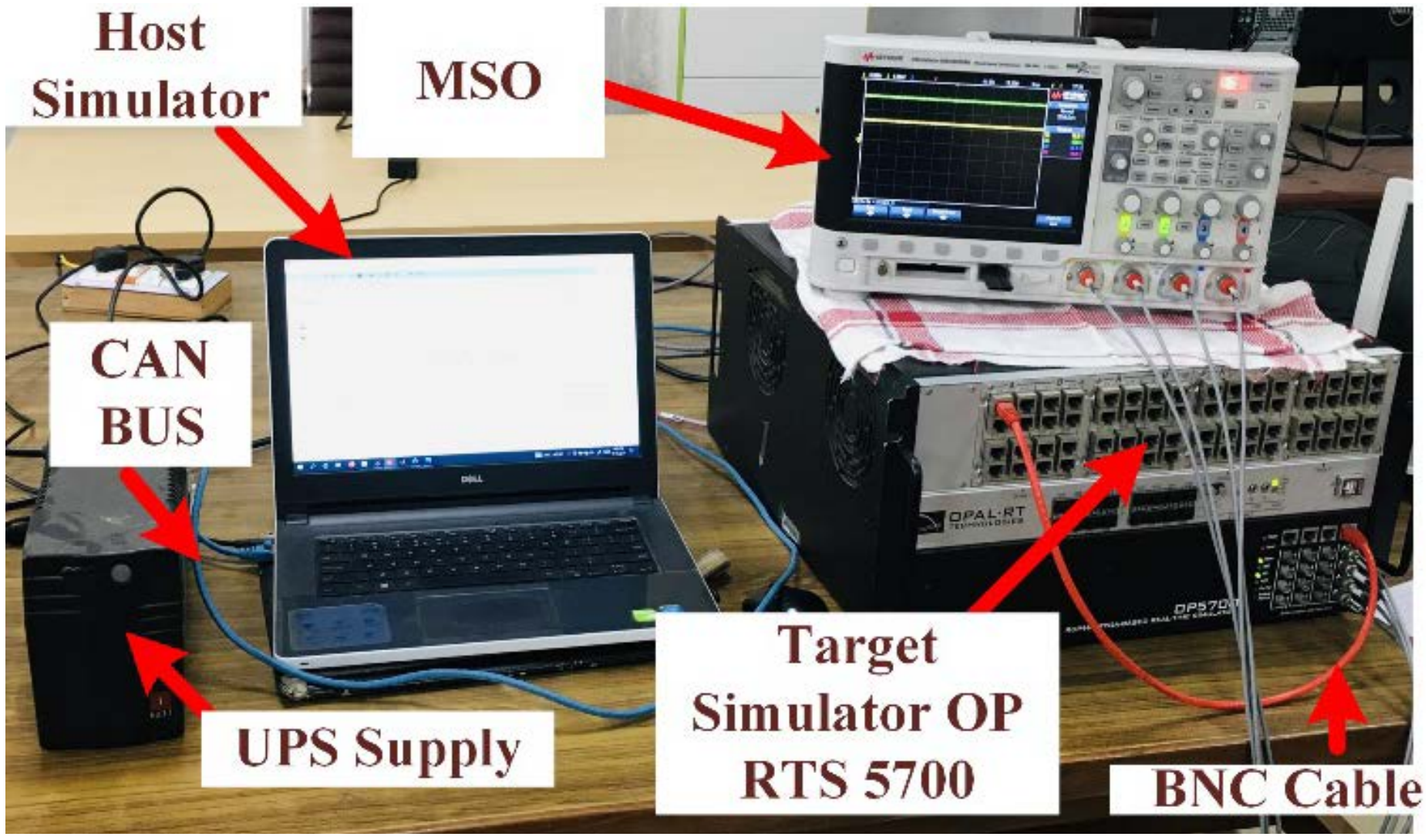

**Fig. 17.** Experiment set-up.

Fig. 18(a) shows the experimental results of constant solar irradiance and SPV power. The fig. 18 (b) shows the nature of load currents having odd harmonics. Moreover, fig.18 (c) shows that at constant solar irradiance grid currents are balanced and sinusoidal. Fig. 18 (d) depicts that the DC link voltage is constant. However, it is noted that the fig. 18 (e) shows that the DC link reference voltage is well settled by applying the ANFIS-based controller. Additionally, it is analyzed that initially, DC link voltage fluctuates from the Vdc reference value. During that initial period, it is noted that the grid currents are distorted in nature which can be analyzed in Fig. 18 (f).

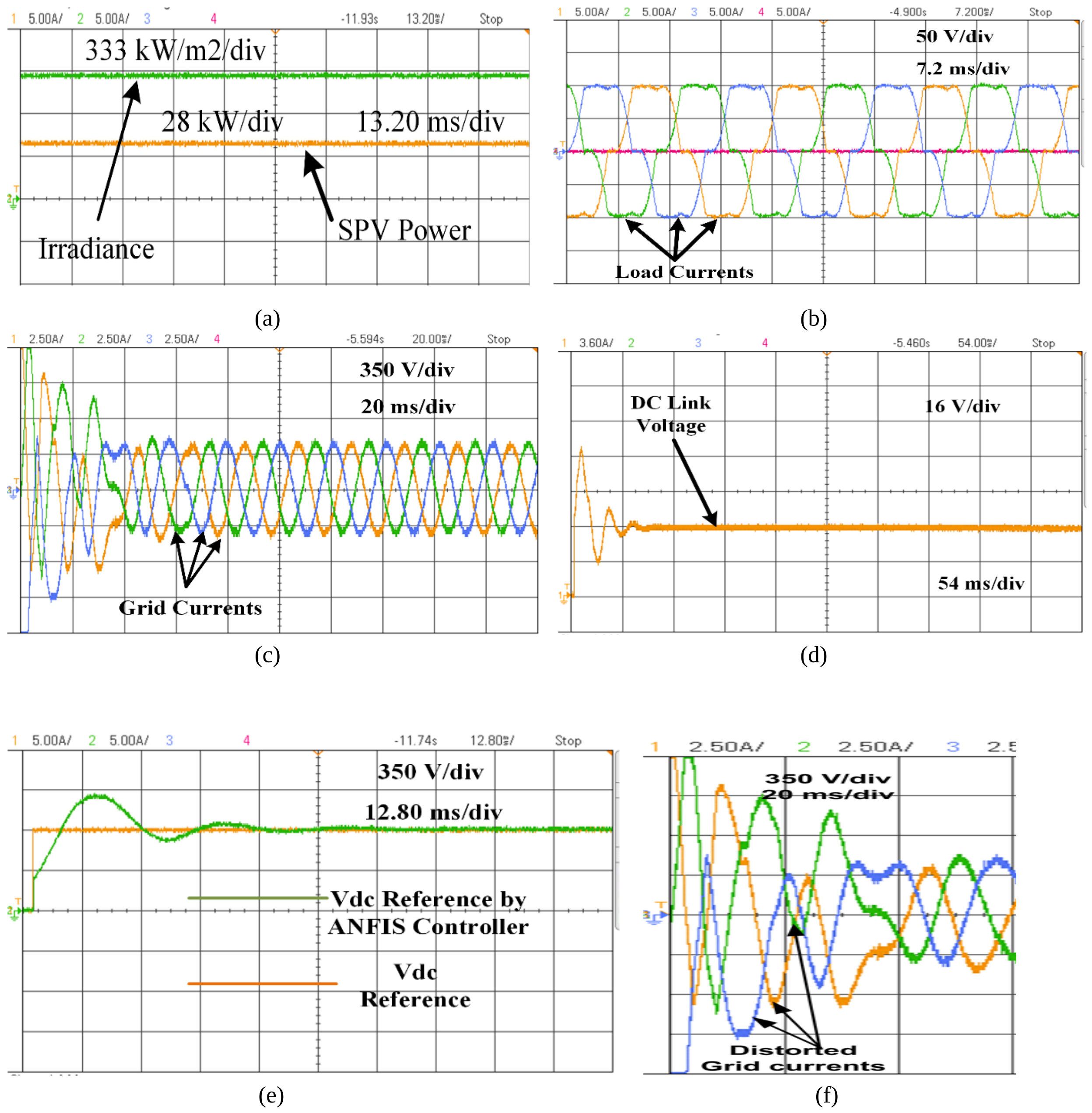

**Fig. 18. (a)** Solar irradiance & Solar PV power **(b)** Load currents **(c)** Grid currents **(d)** DC link voltage **(e)** DC link Voltage by ANFIS controller **(f)** Distorted grid currents

## 7. Conclusions

This paper has discussed and implemented the ANROA-based control algorithms. These control algorithms have been deployed for reactive power compensation and power quality improvement under different circumstances. The proposed technique is the joint execution of ANFIS and ROA techniques and hence it is named the ANROA approach. ANFIS method has been employed to remove maximal power as a PV solar array. The switching pulse for the control of the VSC has been generated using the ROA technique. In addition, the proposed ANROA technique has been implemented in MATLAB / Simulink, and its performance has been assessed using alternative solution techniques. As a result, the suggested Kernel PCA-ESMO technique efficiently lowers power quality issues related to system faults while requiring less computation and reducing the algorithm complexity. The proposed ANROA approach diminishes the THD value for the grid current by 1.26 %. Additionally, the proposed technique is well incorporated in FPGA-based HLL on a real-time simulator for validation.

**Funding:** The work reported herewith has been financially supported by the Spanish Ministerio de Ciencia e Innovación, under the Research Grant RTC2019-007364-3.